\documentclass[runningheads]{llncs}
\usepackage{todonotes}
\usepackage{graphicx}
\usepackage{subcaption}
\usepackage{amsmath}
\usepackage{url}
\usepackage{algorithm}
\usepackage{algpseudocode}
\usepackage{authblk}
\usepackage{amsmath}
\usepackage{amssymb}
\usepackage{booktabs}
\usepackage{hyperref}
\usepackage{enumitem}
\usepackage{comment}

\algrenewcommand\algorithmicforall{\textbf{foreach}}
\algrenewcommand\algorithmicindent{.8em}

\begin{document}

\title{BlockDoor: Blocking Backdoor Based Watermarks in Deep Neural Networks}
\subtitle{Official Work-in-Progress Paper}
\author{Yi Hao Puah \and Anh Tu Ngo \and Nandish Chattopadhyay \and Anupam Chattopadhyay}

\authorrunning{Yi Hao Puah et al.}

\institute{College of Computing and Data Science\\
Nanyang Technological University, Singapore}

\maketitle

\begin{abstract}
Adoption of machine learning models across industries have turned Neural Networks (DNNs) into a prized Intellectual Property (IP), which needs to be protected from being stolen or being used without authorization. This topic gave rise to multiple watermarking schemes, through which, one can establish the ownership of a model. Watermarking using backdooring is the most well established method available in the literature, with specific works demonstrating the difficulty in removing the watermarks, embedded as backdoors within the weights of the network. However, in our work, we have identified a critical flaw in the design of the watermark verification with backdoors, pertaining to the behaviour of the samples of the Trigger Set, which acts as the secret key. In this paper, we present BlockDoor, which is a comprehensive package of techniques that is used as a wrapper to block all three different kinds of Trigger samples, which are used in the literature as means to embed watermarks within the trained neural networks as backdoors. The framework implemented through BlockDoor is able to detect potential Trigger samples, through separate functions for adversarial noise based triggers, out-of-distribution triggers and random label based triggers. Apart from a simple Denial-of-Service for a potential Trigger sample, our approach is also able to modify the Trigger samples for correct machine learning functionality. Extensive evaluation of BlockDoor establishes that it is able to significantly reduce the watermark validation accuracy of the Trigger set by up to $98\%$ without compromising on functionality, delivering up to a less than $1\%$ drop on the clean samples. BlockDoor has been tested on multiple datasets and neural architectures. 


\end{abstract}

\section{Introduction}


For the success of data-driven learning systems, three critical components are essential. First, there must be a large dataset, specifically curated to contain information pertinent to the problem at hand. Second, a neural architecture is required, consisting of a computational graph and weight matrices. Lastly, robust hardware infrastructure is necessary to train the network by tuning the weights or parameters using the available data. The cost of training a neural network can range from a few thousand dollars for models with around a million parameters to over a million dollars for those with more than a billion parameters. Overall, stakeholders must invest in these components to effectively solve tasks using machine learning.

\begin{itemize}
    \item Collecting the relevant data, organizing it, and labeling it, especially for supervised machine learning tasks.
    \item Developing the most appropriate neural architecture for the specific task.
    \item Acquiring advanced hardware infrastructure to train the neural network with the data in order to obtain the learned model.
\end{itemize}


It is understandable that anyone investing in one or more of the previously mentioned components would seek ownership and rights over their assets. It is important to note that these tasks do not need to be performed in one location; they can be executed by different parties collaborating toward a shared goal. Regardless of the method, the trained model, as the outcome of this process, becomes a valuable asset for all parties involved.

\subsection{Motivational Background}

The expansive advancement of deep learning architectures, highlighted by the release of models such as ChatGPT, has underscored the extensive utility of neural network models in our daily lives. Significant resources have been devoted to the development of high-performance models, continually extending their capabilities. 
The limitless applications of these powerful deep learning models have made them economically valuable assets, making them vulnerable to theft, replication, and unauthorized redistribution. As a result, they are increasingly being treated as intellectual property, leading to the adoption of measures to "trademark" them through deep neural network watermarking techniques. The evolution of these techniques began with parameter regularization and has since seen numerous refinements, culminating in the recent introduction of certified watermarking \cite{bansal2022certified}. The initial development of watermarking methods was driven by two essential criteria: robustness, which is the watermark's ability to withstand relentless attacks without being removed, and resilience, which refers to its ability to maintain integrity even under adverse conditions.
Several techniques have been proposed in the past for watermarking neural networks. Notable examples include BlackMarks,  DeepMarks, DeepSigns ~\cite{deepsigns} etc. However, such techniques have proven unreliable due to the simultaneous advancement of attacks by adversaries. Examples of such attack mitigation efforts include DeepInspect ~\cite{deepinspect} and TABOR ~\cite{tabor} which have been successful in defeating them.  

Despite the competition between watermarking techniques and attacks, the most widely accepted and robust mechanism against attacks is watermarking through backdooring ~\cite{mainpaper}. This method leverages the over-parameterized nature of neural networks to embed watermarks within the model. Given its acceptance and widespread use, it is essential to assess its strengths and weaknesses before practical deployment. The backdooring approach takes advantage of known vulnerabilities in neural networks to embed specific watermarks by modifying the model's weights. While effective in both embedding and verifying ownership, this method shares some risks associated with backdoors. Though these vulnerabilities may be less severe in this context, it is still critical to examine potential weaknesses and defenses against adversarial attacks. One relatively unexplored gap is the lack of studies on how to conclusively determine whether a given neural network has been watermarked. Being able to mathematically explain the presence of a watermark in a neural network is crucial for enhancing the security and integrity of digital content, as it provides a foundation for more robust copyright protection and effective tamper detection of intellectual properties. Further details of the watermarking process for neural networks is presented in the Appendix.

\subsection{Contributions}
The primary contributions of this paper are inclusive of, but not limited to the following:
\begin{itemize}
    \item We expose the vulnerabilities of watermarked neural networks which use Trigger sets for backdooring, by detecting Trigger samples and blocking or modifying them to prevent a successful verification of ownership
    \item We have built the BlockDoor framework using a wrapper function around the samples input to the watermarked model, to detect three different kinds of Trigger samples, which correspond to the three ways of Key Generation in Backdooring processes, namely, Adversarial sample based Triggers, Out-of-distribution Triggers and Randomly Labelled Triggers. 
    \item We have performed extensive experiments on standard neural architectures and benchmarking datasets to test the BlockDoor framework, and optimised each component that corresponds to each type of Trigger samples. 
    \item Blockdoor is able to significantly reduce the watermark validation accuracy of the Trigger set by up to $98\%$ without compromising on functionality, delivering up to a less than $1\%$ drop on the clean samples.
\end{itemize}

\section{Breaking Watermarking Schemes}
There are several attack methods that can be used to exploit the vulnerabilities of watermarking schemes and steal models. Two common mechanisms discussed in the literature are evasion attacks ~\cite{evasion} and model modification attacks ~\cite{stealing}. In this work, we present a wrapper based evasion attack that is able to detect all three forms of Trigger samples used in the literature for watermarking with backdooring, and render the watermarked models completely vulnerable to being stolen without authorization.

\subsection{Threat Model}
To validate the ownership of a watermarked model with embedded backdoors, owners will have to pass a trigger set data (image) through their model. The expected result is something that is known only to the owner but not obvious to external users. However, the idea of covering a stolen watermarked model with a wrapper that detects the trigger data, can help prevent such image from ever being passed into the watermarked model or the returned results can be discretely modified. This wrapper solution circumvents the modification of the watermarked model since we encase it in a wrapper that does not directly modify the watermark model. Preserving inherent information of the watermark model. The reduction in watermark accuracy shows that the wrapper models are able to erase some watermark signature without directly modifying the watermarked model.

\subsection*{Neural Network Wrapping}
The idea of developing a wrapper around the potentially watermarked neural network must work for all different types of processes used in generating the embedded watermarks. Presently, three popular techniques for generating trigger data to be watermarked include:
\begin{itemize}
    \item Adversarial Samples
    \item Out of distribution samples
    \item Samples with random labels
\end{itemize}

\begin{figure}[h!]
\centering
\includegraphics[width=\textwidth]{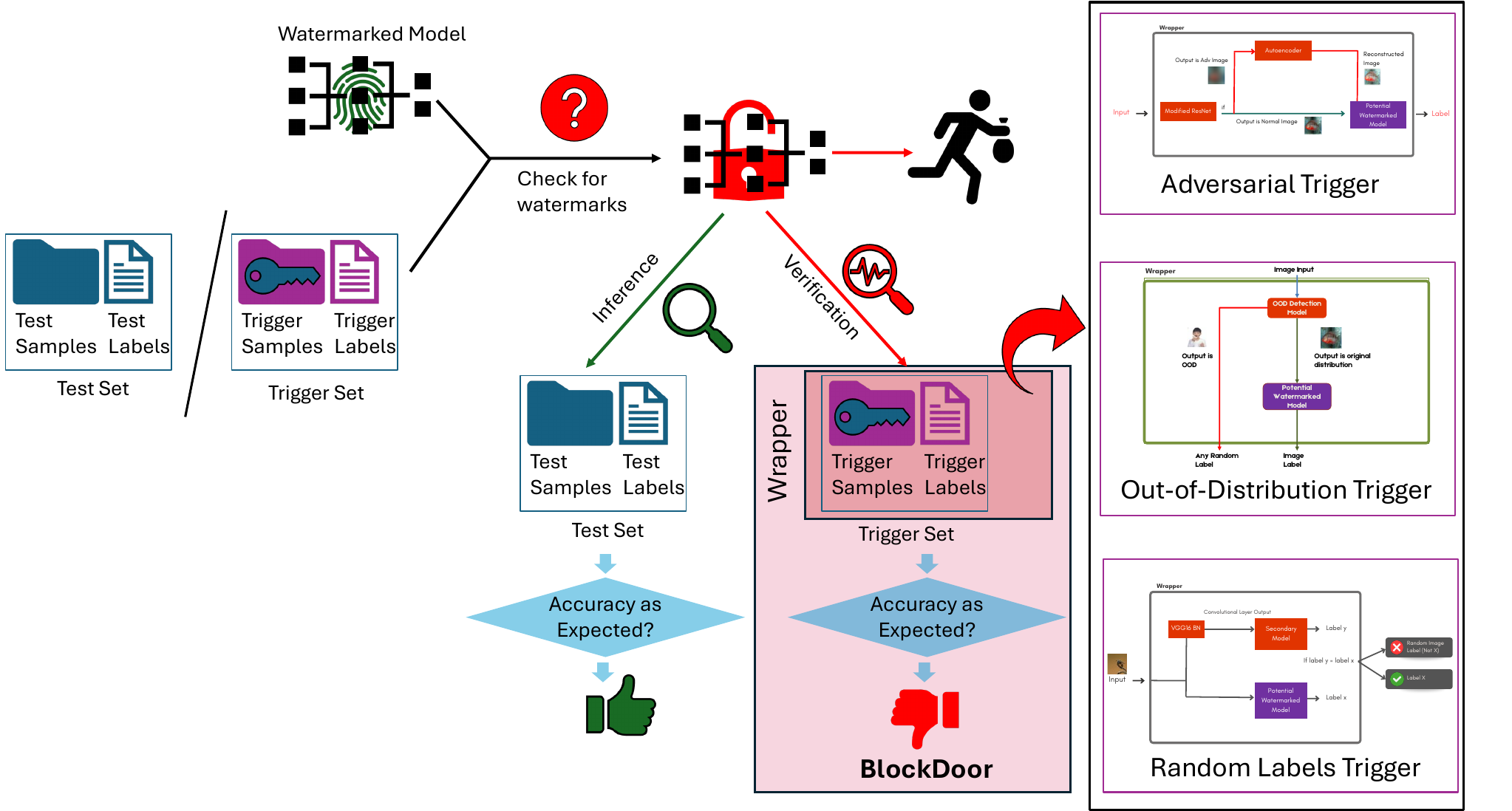}
\caption{BlockDoor in action: Three parallel functionalities to detect and thwart the three different types of Trigger samples that are used for Backdooring processes.}
\label{action}
\end{figure}

BlockDoor is designed to be able to deal with all the three types of Trigger Sets, as shown in Figure \ref{action}. 
Our methodology was based under certain key assumptions and considerations to drove the directions of our proposed techniques. Our key assumptions were:
\begin{itemize}
    \item \textbf{Transparency of Watermarking Scheme:} Following Kerckhoffs' principle, it is assumed that the watermarking scheme is public knowledge. However, the specific trigger data used for embedding the watermark remains confidential. This approach ensures that the security of the watermarking scheme does not rely on the obscurity of the algorithm but rather on the secrecy of the trigger data.
    \item \textbf{Adversarial Access to Model's Dataset:} Considering the scenario where an adversary has access to the model's training dataset. This assumption is realistic in many practical situations where datasets are often publicly available or can be closely approximated.
    \item \textbf{Limited Data for Adversarial Model Training:} Despite having access to the model's dataset, assumptions that the adversary has limited data to train their own model are made. This constraint could arise from various factors, such as computational resources, data access restrictions, or strategic choices by the adversary. In our methodology, this limitation is modelled by randomly sampling only one-third of the original training dataset for all adversarial model training. This assumption helps us evaluate the robustness of our watermarking scheme against adversaries with constrained data resources.
\end{itemize}

\vspace{-5mm}

\subsection{Eliminating Adversarial Backdoor}
Existing research have primarily focused on bolstering robustness against adversarial attacks, as highlighted in the work of Chattopadhyay et al. \cite{chattopadhyay2021robustness}. In contrast, our approach involves adapting this robustness concept by training a dedicated ResNet18 model to differentiate between adversarial and normal image samples. A model specifically trained to recognize adversarial watermarking techniques can effectively identify images altered by such techniques.

\subsubsection{Training Adversarial Detection based Wrapper:}
This method of detecting and thereafter eliminating adversarial noise is a two-step process. First, we modified the ResNet18 architecture to output a binary classification, distinguishing between original and adversarial images. An autoencoder model was trained to minimize the reconstruction loss between an adversarial image and its original counterpart. This process aimed to reverse the effects of the adversarial modifications, effectively 'de-adversarial-attack' the image.

\subsubsection{Two-Step Verification Process:}
The verification process involves two steps. First, an image is passed through the modified ResNet18 model to detect if it is adversarial. If detected, the image is then processed through the autoencoder to reconstruct its original form. This reconstructed image can then be safely fed to the original watermarked model without triggering the watermark detection.

\subsection{Blocking Out-of-Distribution Labelling}
Another popular technique for watermarking a model involves leveraging the use of Out of Distribution (OOD) images, as demonstrated in the work of Wang et al. \cite{wang2022watermarking}. 

\subsubsection{Training OOD Detection Wrapper:}
Out-of-distribution (OOD) data can be effectively detected by converting the problem into a binary classification task, where the original dataset is labeled as positive and randomly sourced data from various datasets is labeled as negative. To validate our hypothesis, we developed a supplementary model specifically trained to distinguish between standard image samples from a dataset and randomly selected images from another dataset. This model is based on the assumption that OOD samples will exhibit significant dissimilarities compared to the original data, making them detectable through binary classification. This methodology is particularly relevant in scenarios where the original trigger set for watermarking may not be directly observable. 

The proposed black box model incorporating our Out of Distribution Detection can be seen as follows:

\begin{algorithm}
\caption{Out of Distribution Watermark Detection}\label{alg:ood}
\begin{algorithmic}
\State \textbf{Data}: Image Samples which may include OOD watermarked images
\State $input \gets image$
\State $label \gets OODModel(input)$
\If{$label$ is 1}
    \State $y \gets potentialWMModel(input)$
\ElsIf{$label$ is 0}
    \State $y \gets randomLabel$
\EndIf
\end{algorithmic}
\end{algorithm}

\vspace{-5mm}

\subsection{Mitigating Random Labelling}
In the paper by Zhao et al. \cite{Zhao_2021}, the authors propose the idea of watermarking Graph Neural Network models using Erdős–Rényi random graphs. In our proposed approach, we simplified this concept by incorporating randomly labelled images as our trigger set data. 
Detection of randomly labeled samples can be effectively achieved using simpler solutions, such as classification models. These models offer a balance between computational efficiency and reasonable accuracy in identifying mislabeled data.

To address the challenge of identifying correct labels without training a model as complex as the watermark model, we propose a two-step classification approach. 

\subsubsection{Training a Wrapper for Randomly Labelled samples:}
The first step involves using a partially trained neural network model to extract features from the images. This model is trained on the correct labels of the original dataset, but not to the point of peak performance. Instead, it learns a representation of the image features, which are more informative than the raw pixel values. we then extract these representations from a hidden layer of the neural network. In the second step, we used a Support Vector Machine (SVM) classifier to classify the extracted features as the original labels. This approach leverages the abstract representations learned by the neural network to make accurate classifications without the need for a fully trained watermark model. This two-step process allows us to efficiently and accurately identify randomly labeled samples, enhancing the robustness of our watermarking technique. This implementation can be visualized in the Figure below:

\vspace{-5mm}


\section{Experimental Results}
The core idea of watermarking neural networks is model and task agnostic, but in this paper, we have demonstrated the watermarking scheme and the wrapper around the Trigger dataset on an image classification task. The primary results related to the breaking of the watermarking verification is presented here, and all other associated results are presented in the Appendix. 

\vspace{-5mm}

\subsection{Results for Adversarial Trigger Detector Wrapper}
Certified watermarking is employed to embed the adversarial samples as the trigger data within a ResNet18 model. To train our adversarial classifier, the original CIFAR-10 data was reclassified as positive labels (1) and the adversarial images as negative labels (0). A ResNet18 model was modified to perform binary classification. As this was a binary classification task, the precision, recall, and F1 score on the test set data was recorded, and noted in the extended results section in the Appendix.
This observation supports our claim which posits that given a simple adversarial watermarking technique such as Fast Gradient Sign Method, it is easy to train a neural network model to learn the pattern of the adversarial disruption.

\vspace{-3mm}

\begin{table}[H]
\centering
\begin{tabular}{l|c|c}
\hline
\textbf{Dataset} & \textbf{Test Accuracy (\%)} & \textbf{Watermark Accuracy (\%)} \\
\hline \hline 
Original Watermarked Model & 84.87 & 100.00 \\
Wrapper Model - ResNet & 84.75 & 12.00 \\
Wrapper Model - ViT & 84.23 & 23.00 \\
\hline
\end{tabular}
\caption{Comparison of test and watermark accuracy between the original watermarked model and the BlockDoor wrapper model - Adversarial Samples as Trigger Set}
\label{table:results}
\end{table}

\subsection{Results for Out-of-Distribution Trigger Detector Wrapper}
Like in the case of detecting adversarial samples, we use a similar technique of re-purposing the primary neural architecture to detect Out-of-Distribution samples.  

\vspace{-5mm}

\subsubsection{Test for segregating within-distribution and out-of-distribution samples:}
Three separate models were trained (MobileNet, ResNet, and VGG11) and their precision, recall, F1 score, and test accuracy across epochs were recorded. 
Our preliminary results indicate that when CIFAR100 is included as a negative label, it is effectively detected as out-of-distribution data. All models achieve optimal performance within a few epochs, with MobileNet demonstrating the most stable F1 score.

\vspace{-4mm}

\begin{table}[H]
\centering
\begin{tabular}{l|c|c}
\hline
\textbf{Dataset} & \textbf{Test Acc (\%)} & \textbf{Watermark Acc (\%)} \\
\hline \hline
Original Watermarked Model & 85.94 & 100.00 \\
Wrapper (MobileNet V2) - Diluted CIFAR100 & 79.83 & 12.00 \\
Wrapper (MobileNet V2) - Excluded CIFAR100 & 77.88 & 83.00 \\
Wrapper (ViT) - Diluted CIFAR100 & 78.53 & 15.00 \\

\hline
\end{tabular}
\caption{Comparison of test and watermark accuracy between the original watermarked model and the wrapper model - Out-of-Distribution samples as Trigger Set }
\label{table:results}
\end{table}

\vspace{-10mm}
We extended this from DNN to Transformer model and achieved a similar performance. Our main concern / future work would be that we still require some OOD data with a similar distribution to the OOD Watermark data to have a stronger presence in erasing its signature, else the erasure is less impactful.

\subsection{Results for Random Labelling Trigger Detector Wrapper}
For our baseline classification model, a VGG16 architecture with Batch Normalization was chosen due to its prior known success in computer vision related task. The hypothesis is that, it is able to extract finer details from the images. As the intended goal of this model was to simply learn features of the original dataset (CIFAR-10), no further fine-tuning is performed.

\vspace{-4mm}

\begin{table}[H]
\centering
\begin{tabular}{l|c|c}
\hline
\textbf{Model} & \textbf{Test Accuracy (\%)} & \textbf{Watermark Accuracy (\%)} \\
\hline \hline
Original Watermarked Model & 85.33 & 100 \\
Wrapper Model - CIFAR10 & 64.45 & 2 \\
Wrapper Model - CINIC10 & 71.11 & 7 \\
\hline
\end{tabular}
\caption{Comparison of test and watermark accuracy between the original watermarked model and the wrapper model - Random Label samples as Trigger Set}
\label{table:results}
\end{table}

\subsection{Key Findings}
The primary take-aways form the experimental analysis include:
\begin{itemize}
    \item BlockDoor's wrapper functions are able to successfully detect all three kinds of Trigger samples that are used in watermarking schemes with backdooring and the corresponding mitigation techniques significantly bring down the accuracy of the watermarked model on the Trigger Set, thereby failing the verification process for establishing ownership. 
    \item BlockDoor satisfies the functionality preserving property, as it is able to reduce the accuracy of the watermarked model on the Trigger Set without bringing down the accuracy of the model on the Test Set. 
    \item For watermarked models which have used Adversarial samples for generating the Trigger Set, BlockDoor is able to bring down the accuracy on the Trigger Set by up to $88\%$ while the accuracy on the Test samples remain within a $1\%$ range. 
    \item For watermarked models which have used Out-of-Distribution samples for generating the Trigger Set, BlockDoor is able to bring down the accuracy on the Trigger Set by up to $88\%$ while the accuracy on the Test samples remain within a best case of $6\%$ range. 
    \item For watermarked models which have used Randomly Labelled samples for generating the Trigger Set, BlockDoor is able to bring down the accuracy on the Trigger Set by up to $98\%$ while the accuracy on the Test samples remain within a best case of $14\%$ range.
\end{itemize}

\section{Conclusions}
Firstly, the experimental results demonstrated that a model trained to recognize a published adversarial technique can effectively neutralize adversarial watermarks. A simple approach was proposed that leverages the transparency of watermarking schemes, and the developed wrapper model has shown the ability to render ineffective most of the trigger data while preserving the high performance of the watermarked model.
Secondly, in addressing out-of-distribution (OOD) data, a promising strategy was presented involving the use of pooled random data to train a basic classifier. This classifier successfully differentiates between the original dataset and OOD data, enabling the creation of a wrapper capable of removing the watermark.
Lastly, for random label watermarks, the proposed method utilizes a partially trained model to extract image features, which are then fed into a machine learning classifier. This approach has proven effective in accurately identifying the original label, enabling random label detection with reasonable accuracy and low computational cost. Overall, the findings suggest that while existing watermarking techniques for neural networks have notable strengths, they also exhibit significant vulnerabilities. The proposed approach exploits these weaknesses to effectively mitigate the watermark from a given model and expose them to be stolen or be used without proper and adequate authorization, which is a severe threat to the IP rights and for proving ownership by stakeholders. 

\subsubsection{Acknowledgement} This work is supported by Nanyang Technological University (NTU)-Desay SV Research Program under Grant 2018-0980.

\bibliographystyle{unsrt}
\bibliography{main}

\newpage

\section*{Appendix}

\section{Background: Watermarking Techniques using Backdooring}

A typical watermarking scheme consists of three key components. Assuming a curated training dataset, denoted as $train\_data$, and a trained neural network model, $M$, the first component is an algorithm to generate a secret key, $m_k$, which serves as the marker to be embedded as the watermark. Alongside this, a corresponding public key, $v_k$, is created for later verification, enabling the detection and validation of the watermark to establish ownership rights. The second component is an algorithm responsible for embedding the watermark into the object, which in this case is the neural network model. Lastly, there is a third algorithm that utilizes both the secret key, $m_k$, for marking and the public key, $v_k$, for verification purposes.

These algorithms can therefore be stated as:
\begin{itemize}
    \item $Key\_Generation()$: Provides the pair of marking and corresponding verification keys $(m_k, v_k)$ 
    \item $Watermark\_Marking(M,m_k)$: Accepts an input neural network model and a secret marking key as parameters $m_k$, returns a watermarked model $\hat{M}$
    \item $Watermark\_Verification(m_k, v_k, M)$: Receives the marking and verification key pair as parameters $(m_k, v_k)$ and the watermarked model $\hat{M}$, returns the output bit $b \in \{0,1\}$ 
\end{itemize}
The effectiveness of the watermarking scheme relies on the proper functioning of all three previously mentioned algorithms $Key\_Generation$, $Watermark\_Marking$ and  $Watermark\_Verification)$ together.

\subsection{Features of Watermarks}
Since the progression into digital watermarks, certain features have been identified to determine its effectiveness. They can be classified in several ways:
\begin{itemize}
    \item \textbf{Robustness}: This refers to the watermark's ability to remain intact and detectable even after the host data has undergone various transformations, such as compression, scaling, cropping, or other forms of manipulation. Robustness is crucial for ensuring that the watermark can effectively protect the copyright and ownership of the digital content.
    \item \textbf{Perceptibility}: This feature measures the visibility of the watermark in the host data. Ideally, a watermark should be imperceptible to maintain the quality and usability of the original data while still being detectable through specific algorithms or techniques. Balancing perceptibility and robustness is a key challenge in watermarking design
    \item \textbf{Capacity}: This refers to the volume of information that can be incorporated within the watermark. Higher capacity allows for more data to be stored, such as copyright information, authentication codes, or metadata. However, increasing capacity often requires trade-offs with perceptibility and robustness.
    \item \textbf{Security}: The security of a watermark is its ability to resist unauthorized detection, removal, or alteration. This is achieved through cryptographic techniques, watermarking key management, and embedding strategies that make the watermark difficult to tamper with or replicate without authorization. 
\end{itemize}

\subsection{Watermarking for Neural Networks}
\begin{figure}[h!]
\centering
\includegraphics[width=\textwidth]{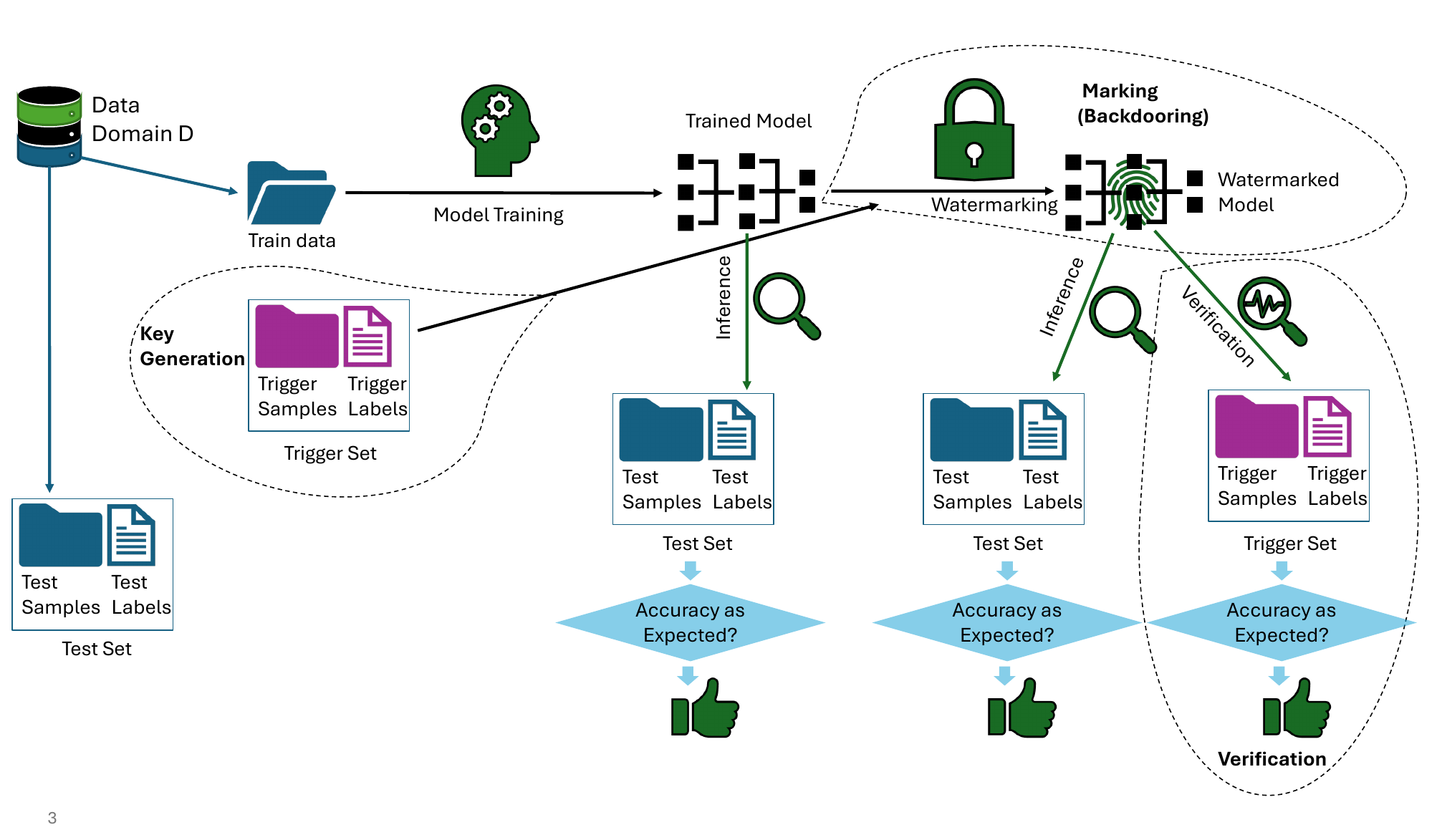}
\caption{Schematic diagram of the watermarking scheme using Backdooring which uses Trigger samples for Verification}
\label{schematic_1}
\end{figure}

An abstract visualization of the watermarking process is shown in Figure \ref{schematic_1}.
The specific steps of the process are discussed here:
\begin{itemize}
    \item \textbf{Selection of Watermark:} The first step involves choosing a suitable watermark. This could be a specific pattern, a set of weights, or a unique configuration that can be embedded into the neural network. The watermark should be distinctive enough to assert ownership but should not significantly impact the performance of the model.
    \item \textbf{Embedding the Watermark:} Once the watermark is selected, it is embedded into the neural network. This can be done in various ways, such as modifying the weights of the network, inserting specific neurons or layers that encode the watermark, or altering the network's architecture to incorporate the watermark. The embedding method should ensure that the watermark is integrated seamlessly without degrading the model's performance.
    \item \textbf{Extraction and Verification:} After the watermark is embedded, it is essential to have a mechanism for extracting and verifying the watermark to prove ownership. This involves developing an algorithm or technique that can detect the presence of the watermark in the neural network and confirm its authenticity.
    \item \textbf{Robustness and Security:} The watermarking process should also consider the robustness and security of the watermark. It should be resistant to common attacks such as fine-tuning, model compression, and adversarial attacks that might attempt to remove or alter the watermark. Additionally, the watermark should be secure enough to prevent unauthorized extraction or duplication.
\end{itemize}

\begin{figure}[h!]
\centering
\includegraphics[width=\textwidth]{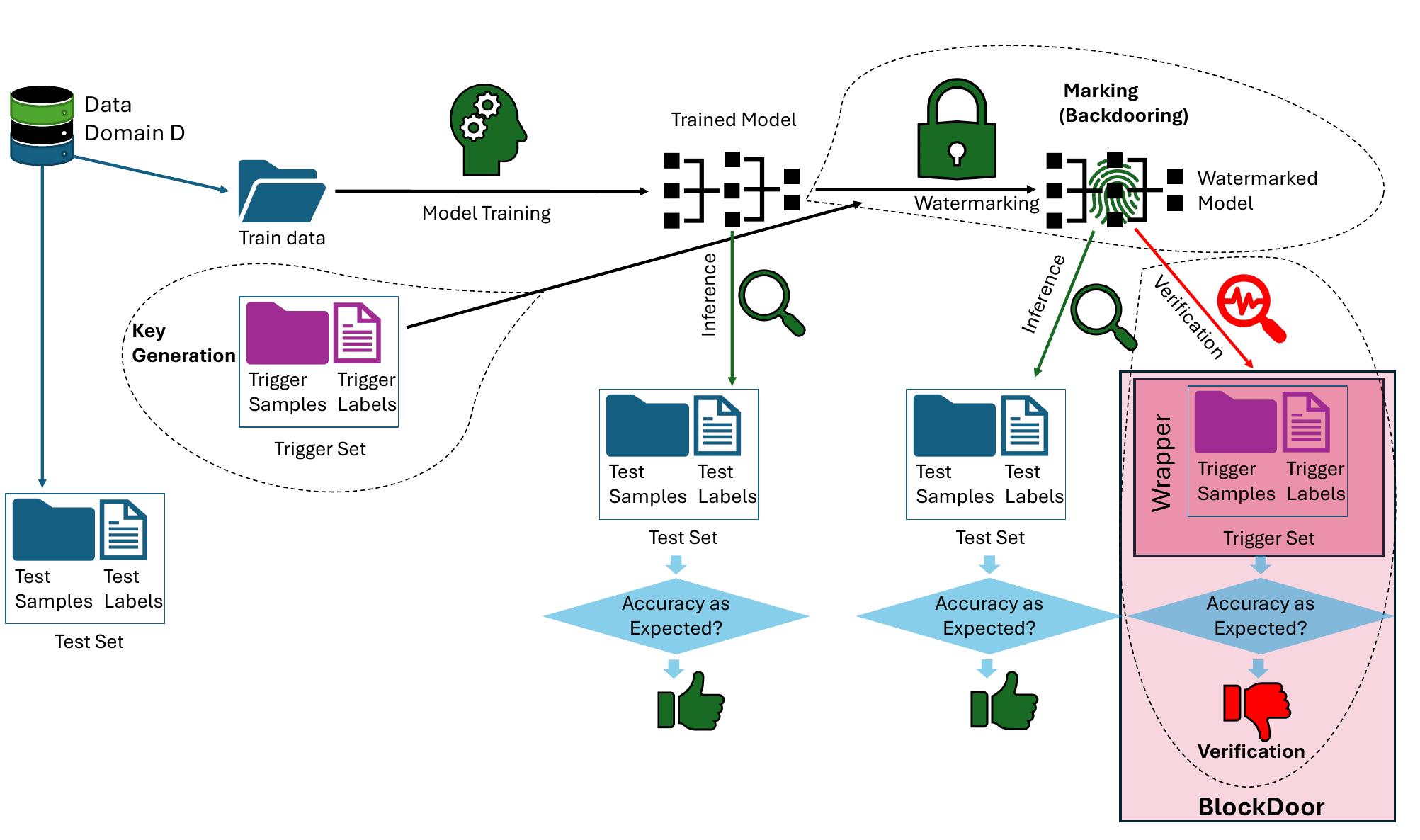}
\caption{Schematic Diagram of BlockDoor: Blocking Backdoor Based Watermarks in Deep Neural Networks}
\end{figure}

\subsection{Models and Datasets}
The underlying model which is being watermarked, is what we refer to as the primary model. For this, we have used three different neural architectures. 

Firstly, ResNet (Residual Network) architecture is employed, which has proven effective in addressing the vanishing gradient problem and enabling the training of very deep neural networks. There have been variations in the code of ResNet, but references were made based on Sasha et al, 2016 \cite{targ2016resnet} implementation of ResNet. In addition to ResNet, the VGG (Visual Geometry Group) network architecture was also utilized for the image classification task. Specifically, a VGG16 architecture was employed with Batch Normalization, which is an enhancement over the standard VGG16 model. This architecture was first introduced by Simonyan and Zisserman in their 2014 paper \cite{simonyan2014very} and has since become a popular choice for various computer vision tasks. Thirdly, MobileNet is used, which is a lightweight convolutional neural network architecture designed specifically for mobile and embedded devices. It was introduced by Howard et al. in their 2017 paper \cite{howard2017mobilenets}. The key feature of MobileNet is the use of depthwise separable convolutions, which significantly reduces the number of parameters and computational complexity compared to traditional convolutional layers. 

Additionally, for the design of the wrappers itself, we have made use of vision transformers and autoencoders, as explained earlier. The architecture of an autoencoder consists of two main components: an encoder and a decoder. The encoder compresses the input data into a latent-space representation, while the decoder reconstructs the input data from the latent space. Mathematically, the encoder and decoder can be represented as functions \(f\) and \(g\), respectively, where: $h = f(x) $ ,  $ \hat{x} = g(h) $ where \(x\) is the input data, \(h\) is the encoded representation, and \(\hat{x}\) is the reconstructed data. The autoencoder is trained to minimize the reconstruction error, typically measured by the mean squared error (MSE) between the input \(x\) and the reconstructed output \(\hat{x}\).

\subsection{Datasets}
The CIFAR-10 dataset is commonly used for evaluating image classification models and serves as a standard benchmark for comparing the performance of different algorithms. \cite{cifar}. The CIFAR-100 dataset is an extension of the CIFAR-10 dataset, consisting of 100 classes containing 600 images of dimensions 32 × 32 each. The classes range from rockets to fishes to flowers. The CINIC-10 dataset \cite{darlow2018cinic} is an extension of the CIFAR-10 dataset designed to improve the evaluation of machine learning models on image classification tasks. It includes 270,000 images across 10 classes, which are a combination of CIFAR-10 images and additional images from the ImageNet dataset. This extended dataset aims to provide a more diverse and challenging set of images, helping to assess model performance more robustly and address issues related to overfitting and generalization. The SVHN dataset is a real-world image dataset obtained from house numbers in Google Street View images. It consists of over 600,000 digit images, coming in two formats. 

\section{Results}
\subsection{Eliminating Adversarial Backdoor}

\begin{figure}[h!]
    \centering
    \includegraphics[width=\textwidth]{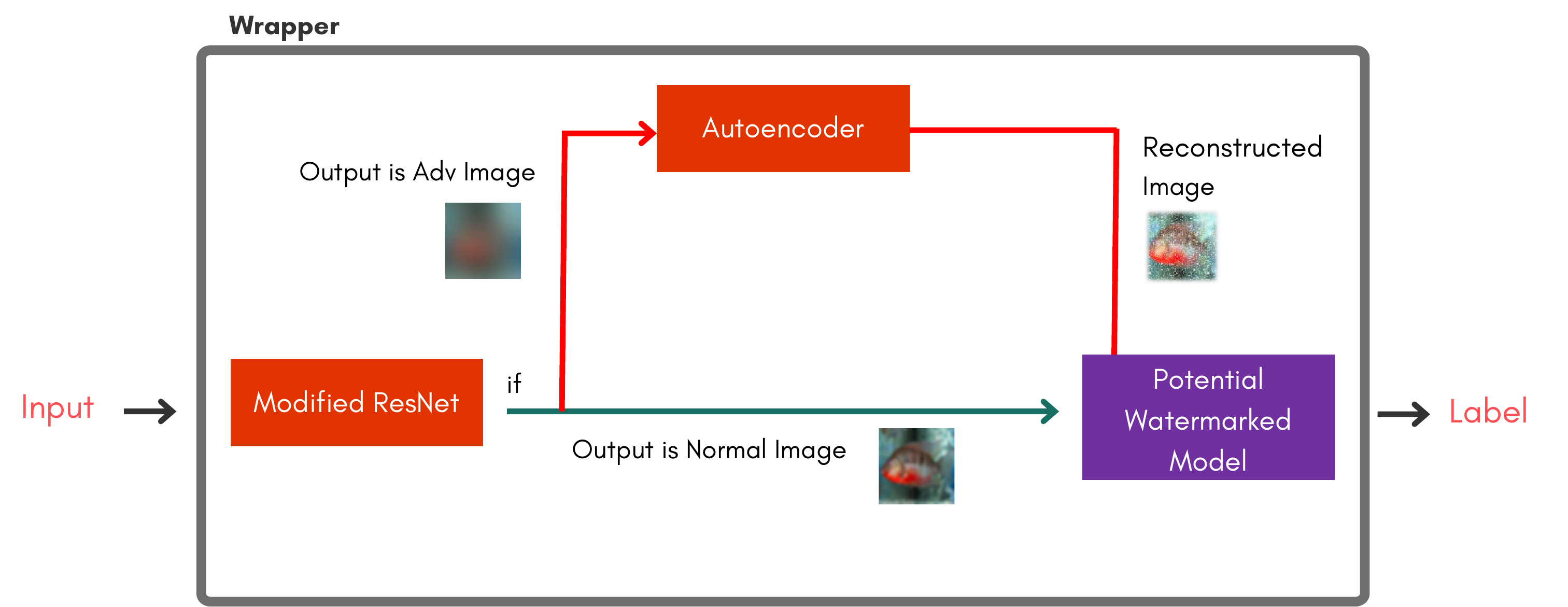}
    \caption{BlockDoor architecture with Wrapper Function for detecting and eliminating Trigger Samples containing Adversarial Noise. }
\end{figure}

The ResNet-18 model was trained with the following parameters: \textbf{Loss Function:} Cross Entropy Loss, \textbf{Optimizer:} Adam Optimizer, \textbf{Learning Rate:} 0.001, \textbf{Batch Size:} 128, \textbf{Training Epochs:} 20. 
To train our adversarial classifier, the original CIFAR-10 data was reclassified as positive labels (1) and the adversarial images as negative labels (0). A ResNet18 model was modified to perform binary classification. 
The train - test accuracy during the training was recorded

\begin{figure}[h!]
\centering
\includegraphics[width=0.8\textwidth]{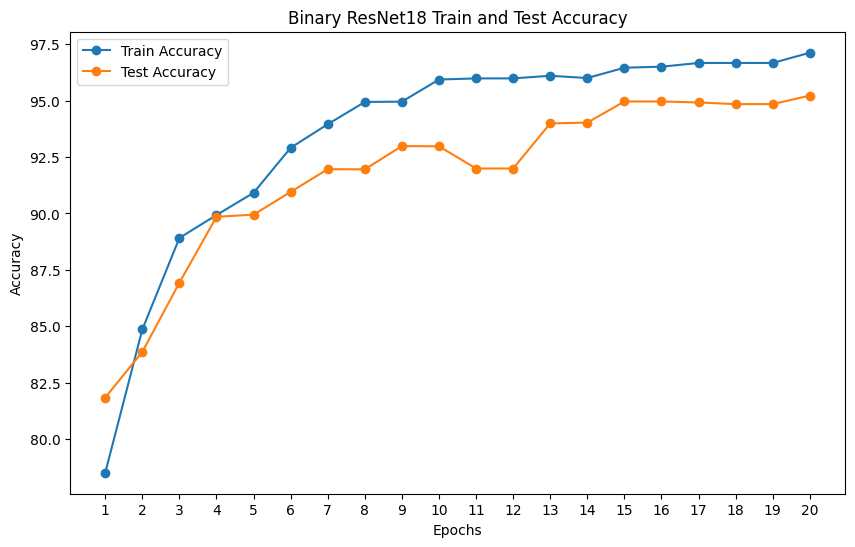}
\caption{ResNet18 Train-Test accuracy}
\end{figure}

Additionally, the experiment was repeated with a Vision Transformer (ViT) model to show generalisability. The Vision Transformer (ViT) model \cite{yuan2021tokens} is a deep learning architecture that applies transformer principles, originally designed for natural language processing, to image analysis. Unlike traditional convolutional neural networks (CNNs), ViT processes images by dividing them into fixed-size patches, linearly embedding these patches, and then applying a transformer encoder to capture global relationships within the image. This approach has demonstrated significant performance improvements on various image classification benchmarks, showcasing the transformer model's effectiveness in vision tasks. The results of the various models are documented as shown below:

\begin{table}[H]
\centering
\begin{tabular}{l|c|c}
\hline
\textbf{Dataset} & \textbf{Test Acc (\%)} & \textbf{Watermark Acc (\%)} \\
\hline \hline
Original Watermarked Model & 84.87 & 100 \\
Wrapper Model - ResNet & 84.87 & 11.2 \\
Wrapper Model - ViT & 84.23 & 22.8 \\
\hline
\end{tabular}
\caption{Comparison of test and watermark accuracy between the original watermarked
model and the BlockDoor wrapper model - Adversarial Samples as
Trigger Set}
\label{table:adv}
\end{table}

\subsection{Blocking Out-of-Distribution Labelling}

\begin{figure}[h!]
    \centering
    \includegraphics[width=0.8\textwidth]{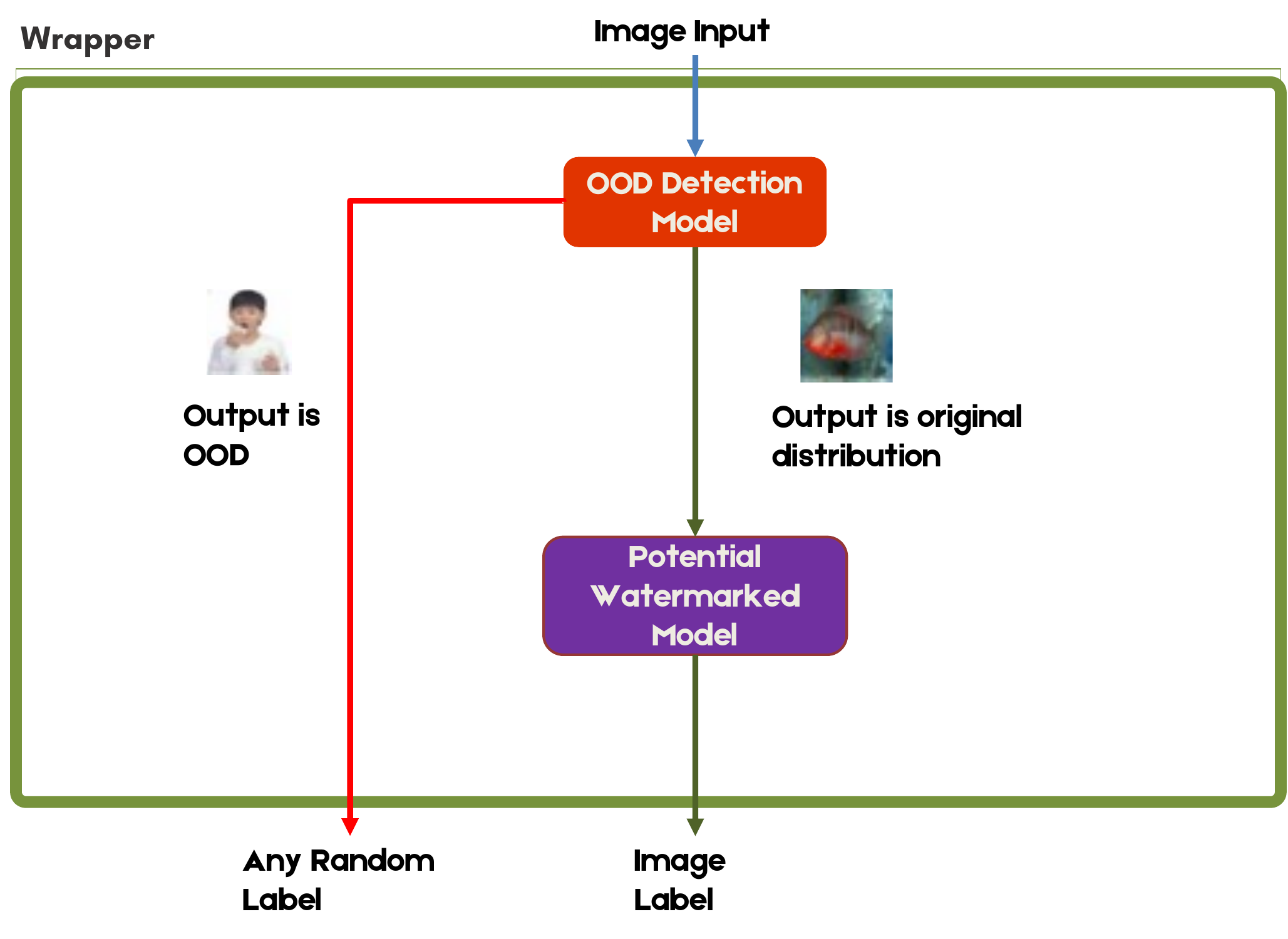}
    \caption{BlockDoor architecture with Wrapper Function for detecting and blocking Out-of-Distribution samples used as Trigger Samples}
\end{figure}

The following settings are used for the same: \textbf{Loss Function:} Cross Entropy Loss, \textbf{Optimizer:} Adam Optimizer, \textbf{Learning Rate:} 0.01, \textbf{Batch Size:} 128, \textbf{Training Epochs:} 20.

\subsubsection{Negative Labels for Out-of-Distribution samples:}
The experiments were carried out with a diluted dataset and recorded the same metrics. The figure below showcases the classification F1 score for out-of-distribution data.

\begin{figure}[h!]
\centering
\includegraphics[width=0.8\textwidth]{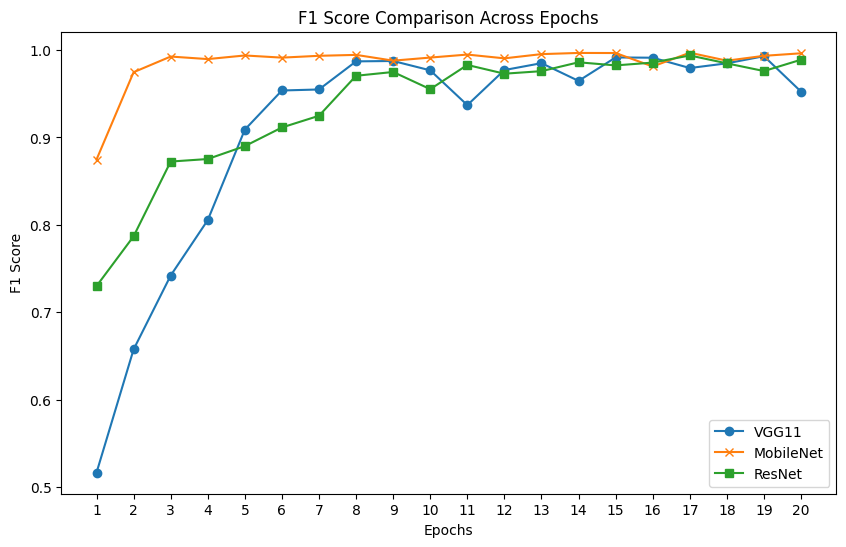}
\caption{Detection of Out-of-Distribution samples (with negative labels) re-purposing the primary neural architectures.}
\end{figure}

\subsubsection{Extending to unseen/excluded Out-of-Distribution samples:}
Based on the previous results, it was determined that MobileNet was the most suitable model for identifying out-of-distribution data. Hence, training was focused solely on the MobileNet model, on a dataset with CIFAR100 completely excluded from the negative label. For the purpose of fine-tuning this model, additional features were added during the training step, such as EarlyStopping (5) and a Learning Rate Scheduler based on the test loss. 

\begin{figure}[h!]
\centering
\includegraphics[width=0.9\textwidth]{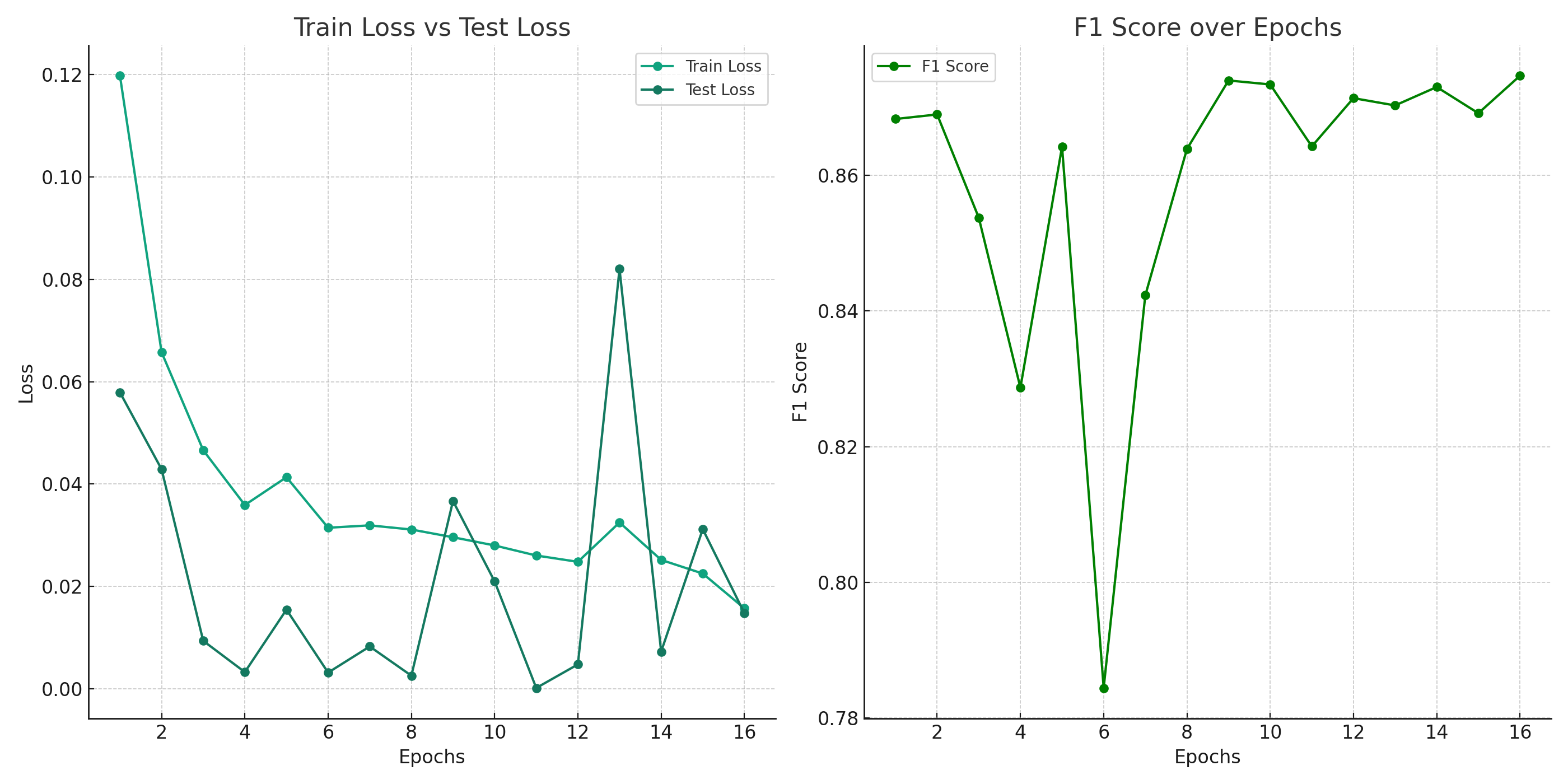}
\caption{Accuracy of Detection of Out-of-Distribution samples when the watermarked  model is trained on the CIFAR-10 dataset and the Trigger samples are from the CIFAR-100 dataset.}
\end{figure}

\subsection{Mitigating Random Labelling}

\begin{figure}[h!]
    \centering
    \includegraphics[width=0.8\textwidth]{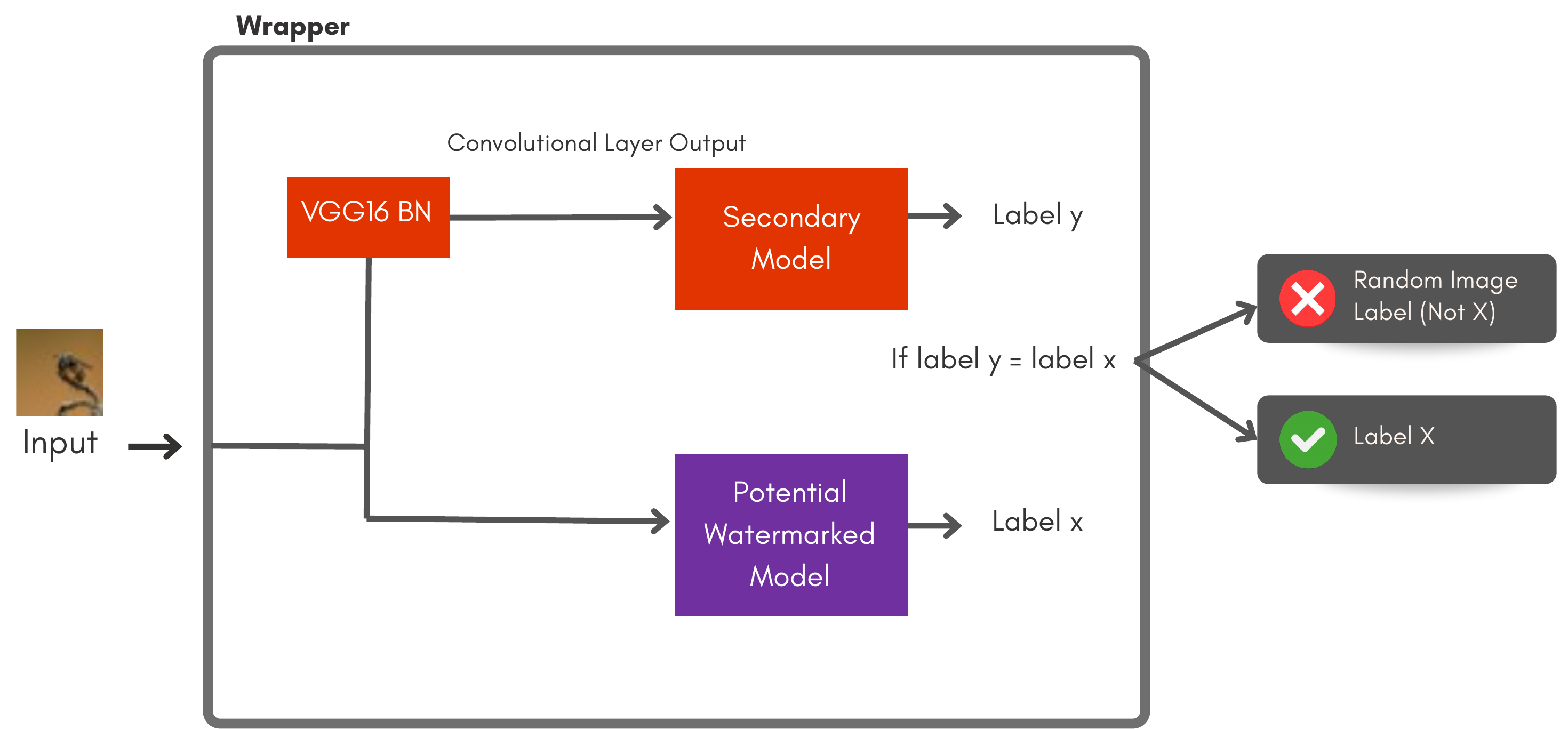}
    \caption{BlockDoor architecture with Wrapper Function for detecting and mitigating Randomly Labelled Samples used as Trigger Samples}
\end{figure}

This model has the following parameters: \textbf{Loss Function:} Cross Entropy Loss, \textbf{Optimizer:} Adam Optimizer, \textbf{Learning Rate:} 0.01, \textbf{Batch Size:} 64, \textbf{Training Epochs:} 50.

\subsubsection{Training the alternate model:}
After training the VGG16 neural network, features were extracted from the 2nd last Convolutional 2D layer. These extracted features were then used as input for simpler machine learning models to evaluate their classification performance. Specifically, Support Vector Machine (SVM) and K-Means clustering algorithms were employed. To ensure a proper mapping of label inputs for the K-Means clustering, the Hungarian algorithm is applied before computing the cluster label accuracy. The classification performance was assessed using 5-Fold Cross-validation accuracy. The results of the classification performance for the SVM and K-Means models are summarized in the following table:

\begin{table}[H]
\centering
\begin{tabular}{l|c|c}
\hline
\textbf{Model} & \textbf{Accuracy} & \textbf{Standard Deviation (\%)} \\
\hline \hline 
K-Means & 0.649 & $\pm$ 1.24 \\
SVM & 0.7811 & $\pm$ 1.44 \\
\hline
\end{tabular}
\caption{Classification performance of SVM and K-Means models using 5-Fold Cross-validation.}
\label{tab:classification_results}
\end{table}

Given the promising results from the SVM model, further exploration is made to examine the potential of enhancing its performance by reducing the dimensionality of the input features. Principle Component Analysis (PCA) is applied to the extracted features from the convolutional layer, which initially had 2049 dimensions. Our goal was to investigate whether reducing the number of features could optimize the classification performance.
The results of applying PCA with various numbers of components are summarized in the following table:

\begin{table}[H]
\centering
\begin{tabular}{l|c|c}
\hline
\textbf{n\_components} & \textbf{Cross-Validation Acc (\%)} & \textbf{Cross-Validation Std. Dev. (\%)} \\
\hline \hline
Unmodified & 78.11 & $\pm$ 1.39 \\
0.95 & 80.11 & $\pm$ 1.48 \\
0.90 & 80.03 & $\pm$ 1.39 \\
0.85 & 80.03 & $\pm$ 1.39 \\
\hline
\end{tabular}
\caption{Performance of the SVM model with PCA-applied feature reduction.}
\label{table:pca_results}
\end{table}
The results indicate that reducing the dimensionality of the feature space through PCA can lead to improved classification performance. Specifically, retaining 95\% of the variance in the data (n\_components = 0.95) resulted in the highest cross-validation accuracy.

\end{document}